# COMPUTATIONAL FLUID DYNAMICS AS AN EMERGING SUPPORTING CLINICAL TOOL: REVIEW ON HUMAN AIRWAYS

Rajneesh Kashyap, R. Thundil Karuppa Raj

School of mechanical and building Science, VIT University,Vellore-632014,India

**Correspondence:**

Prof. R. Thundil Karuppa Raj (PhD)
Professor
School of Mechanical Engineering, VIT University, Vellore – 632014, India
email:thundilr@gmail.com

**Authorship and Contributorship:**

**R. Thundil Karuppa Raj**:  Designed the review study, Data analysis and interpretation, Edited and approved the final manuscript.
**Rajneesh Kashyap**: Contributed to literature search, Data acquisition, Drafted the manuscript

**Conflict of Interest:**
There are no conflict of interest in connection with this article.

**Abstract:**

*Objectives:* The main objective of this review article is to evaluate the usability of Computational Fluid Dynamics (CFD) as a supporting clinical tool for respiratory system.

*Data Source:* The English articles referred for this review paper were identified from various International peer reviewed journals indexed in Science citation index.

*Study Selection:* 26 high quality articles most relevant to the highlighted topic which were published in last fifteen years were selected from almost 120 articles.

*Results:* The analysis done and the outcome obtained by this computational method is as accurate as Spirometry and Pulmonary function test (PFT) result. CFD can be very useful in the cases where patents is unable to perform PFT. Pressure drop, Velocity profile, Wall shear stress & other flow parameter, respiratory resistance, Pattern of drug deposition, Particles transport/deposition, etc. had also been predicted accurately using CFD.

The effect of tracheal stenosis on the flow parameters has been predicted. The size and location of tracheal stenosis has also been correlated with breathing difficulties. The distribution of air in various lobes of the lungs can be accurately predicted with CFD tool.

*Conclusion:* Virtual surgery is eventually possible by using CFD after further research with validation. With the help of this multi - disciplinary and efficient tool we can obtain accurate result while reducing cost and time.

**Keywords:**

**Introduction:**

The working fluid air, through the human being has complicated mechanism which is very difficult to examine. Studies to learn the behavior of the fluid is laborious and expensive process. Computational Fluid Dynamics (CFD) compliments the analytical and experimental method by presenting a different and profitable means of simulating real time fluid flow, especially in human body fluids. CFD has the ability to simulate flow conditions that are not attainable while performing experimental tests that are difficult to be experimentally simulated. The passage of working fluid air through nostrils, trachea, bronchi and lungs can be efficiently and effectively predicted with the help of CFD tool and the proportion of air distribution can also be estimated accurately. For patients who cannot be assessed by the pulmonary function test (PFT), the CFD technique seems to be most promising tool and is essential to diagnose the air pathway. CFD is a subsidiary of fluid mechanics, branch of science that makes use of numerical analysis to find a solution to problems that involve flow of fluids with the help of modern high end computers. CFD solves highly non-linear partial differential governing flow equations namely conservation of mass, momentum and energy. The turbulence induced in the flow path can be modeled using turbulence mode of closure. Calculations essential to simulate the interaction of fluids with the surfaces characterized by boundary conditions, are carried out using a computer.

Anticipating the flow of air in the tracheal path and the distribution of air in bronchi and lobes of human lungs is difficult for a variety of reasons. These also include the drawbacks of current imaging technologies, deviations in lung morphology specific to the patient, and alterations in the geometry of the airway, caused during the breathing cycle, and also, due to the extreme size and multi-scale quality of the airway geometry (1). Exact information regarding the flow of air and deposition of particle is extremely essential in order to understand the impact of inhaled particulates to human health. These comprise potentially hazardous substances such as, pollutants to the environment, in addition to aerosolized medicines used in pulmonary drug delivery. The capacity to competently model the flow of air in the lungs would provide a beneficial substitute to expensive and potentially hazardous *in vivo* experimental methods (2).

CFD is competent enough to study the real-time dynamics of the airflow and the changes involved in the obstruction of the central airway, before and after the implementation of the airway stent. The numerical results obtained through CFD, correlate well with those obtained using PFT's, which imply that CFD can potentially become a useful tool for clinical prognosis prediction, in order to study lung cancer patients, who have deteriorating condition and for whom PFT's cannot be successfully applied (3). CFD has been used to describe the flow of fluid in the human airway models. It has achieved significant attention, both in the field of engineering as well as medical sciences, primarily due to its non-invasive nature. It has the potential in determining the characteristics of the flowing fluid under transient conditions where one or even more input flow variables changes with time (4). It also allows the analysis of air flow path for different flow variables and fluid forces to very minute detail (4).

CFD analysis constitute of three stages: pre-processing, processing and post processing as shown in Figure 1. Graphical representation and modelling is the most primary step which in this field is done by stacking slices of CT image together, extracting the area of interest and remodeling with the help of image processing and 3D modelling software. The next stage of preprocessing involves disintegration into nodal points and forming the mesh of discrete small finite volumes. Final step of preprocessing is feeding the solver with boundary condition such as mass flow rate, viscosity of air, body temperature, air pressure etc. Processing stage is where the relevant Governing equation required are chosen to be solved at all the nodal points. Conservation of mass, momentum and energy are three essential governing equation, as shown below eq. 1 to 5, to be solved followed by turbulence equation if feasible. These equation are iterated till the convergence criteria is fulfilled. Results from the previous stage are converted to interpretable form in post processing.

$$\textit{Conservation of mass: } \nabla\left(\rho\,\vec{V}\right) = 0 \tag{1}$$

$$\textit{x - Momentum: } \nabla\left(\rho u\,\vec{V}\right) = -\frac{\partial p}{\partial x} + \frac{\partial \tau_{xx}}{\partial x} + \frac{\partial \tau_{yx}}{\partial y} + \frac{\partial \tau_{zx}}{\partial z} \tag{2}$$

$$\textit{y - Momentum: } \nabla\left(\rho u\,\vec{V}\right) = -\frac{\partial p}{\partial y} + \frac{\partial \tau_{xy}}{\partial x} + \frac{\partial \tau_{yy}}{\partial y} + \frac{\partial \tau_{zy}}{\partial z} \tag{3}$$

$$\textit{z - Momentum: } \nabla\left(\rho u\,\vec{V}\right) = -\frac{\partial p}{\partial z} + \frac{\partial \tau_{xz}}{\partial x} + \frac{\partial \tau_{yz}}{\partial y} + \frac{\partial \tau_{zz}}{\partial z} \tag{4}$$

$$\textit{Energy: } \nabla\left(\rho e\,\vec{V}\right) = -p\,\nabla\,\vec{V} + \nabla\,(k\nabla\,T) + q + \emptyset \tag{5}$$

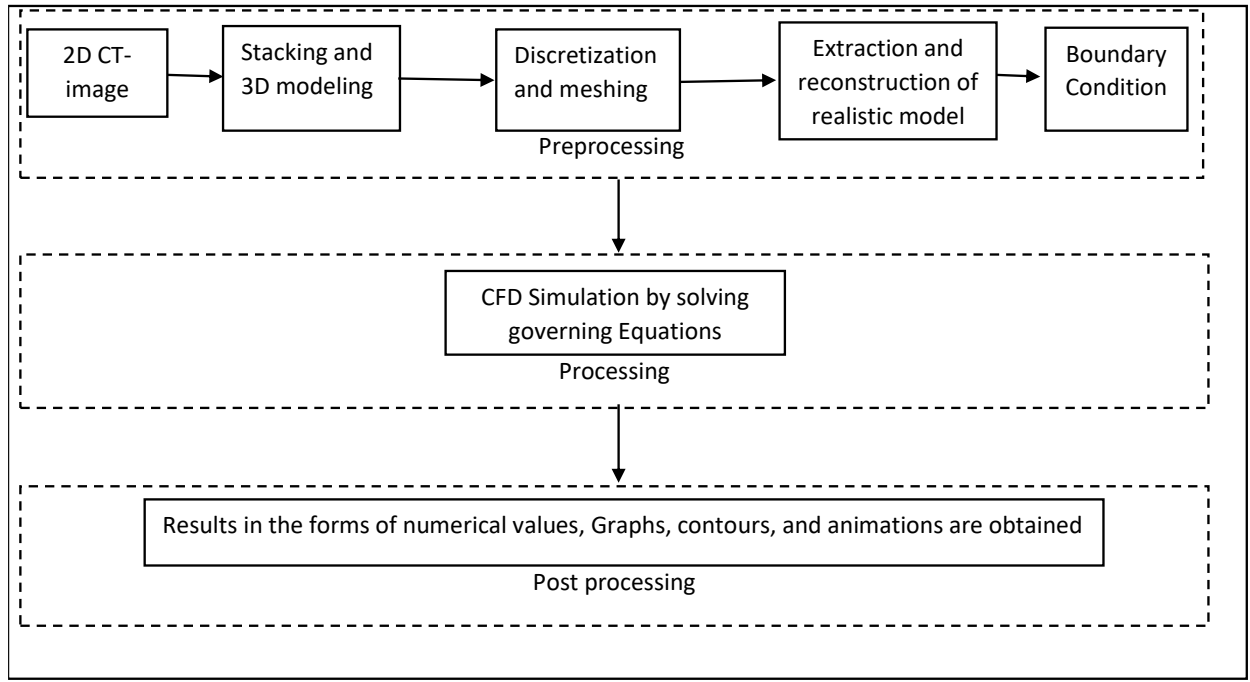

Figure 1 – Process flow diagram

**Modelling Human airways for CFD analysis:**

Computed tomography (CT) is an accepted imaging method which is frequently required for analysis and estimation of diseases in the lungs. It also helps physicians to detect and diagnose pathological conditions.

A noteworthy study on automated segmentation of human airways was done by Pu *et al*. (5) .They came up with a computerized scheme to automatically segment the 3-D human airway tree as shown by the CT scan images. This method has the advantage of principal curvature as well as principal directions in understanding the difference between airways and the other tissues in the surrounding geometric area. For this study they collected and examined 75 cases of chest CT scan with different slice thickness on 45 Subject and also used 20 case studies which were publicly available for making algorithm which can easily segment human airways tree. Dong *et al*. (6) constructed the human upper airways using the basic 2-D CT scan images. The images initially processed using MIMICS in order to obtain coarse 3-D model of upper respiratory airway geometries for each patient. The coarser models were then produced from scans in a four step mechanism which include importing the image files and setting the parameters for calibration, breaking down the images & preparing the boundaries,  rendering surface & exporting coarse 3-D models, and finally implementing the smooth parameters and smoothening the coarse models. An anatomically rational three-dimensional (3D) model was obtained by Johari *et al*.(7) from digital imaging and communications in medicine (DICOM) format CT scan images of a 50 year old Asian female. The 3D model was made by using MIMICS software, which is generally geometry & file conversion software. They also observed that there was no abnormality found throughout the scan. Thus, that was concluded to be a good portrayal of healthy human airways

According to Nardelli *et al*.(8),  in certain situations having an airway distribution method, which allows for reconstruction of the airway from chest CT images can aid massively in the study of diseases in the lungs. According to them lots of efforts have been put into the development of airway segmentation algorithms. However, methods are generally not optimized to be dependable across different CT scan parameters. Their semi-automatic algorithm was found to be highly stable compared to others. They concluded that by using their algorithm in 3D slicer platform, an open source software, tracheal and bronchial anatomy can easily be segmented.

To construct model from CT scan image, many competitive software have been developed and each has their own features. Gorgi *et al*.(4) have used 3D doctor software to construct human airways which was then reconstructed and modified using CATIA solid modelling. Shih *et al*.(9) have used combination of Amira and ANSYS software to re-construct and meshing the model for analysis. Ultimately these researchers obtained realistic 3D model of human airways.

**Boundary condition:**

The decisive step to obtain analytical results is to provide boundary condition which is in sync with the human breathing system. Physical activity dominates the working of human breathing system.  Different mass flow rate, inspiratory and expiratory air flow, body temperature, density & viscosity of air at inspiratory and expiratory phase etc. are few parameters constitutes boundary condition.

The inlet boundary conditions of velocity have been applied on the basis of the internal flow rate studied by Nithiarasu *et al*. (10). They considered Normal no-slip boundary conditions on the solid walls. An extremely small value of the turbulence parameter was defined at the inlet. This parameter was given an assumed value of zero on the walls. While performing the calculations, the variation between the inlet and outlet mass flow rate was examined and brought down to a value much below the prescribed value of 5%. Since the explicit scheme was used, the outlet pressure was automatically maintained at a value of zero, which was the initial condition. By using the commercial tool ANSYS, Wang *et al*.(11) Simulated the inspiratory as well as the expiratory flow of air though the upper airway model. According to earlier studies, it was reported that the turbulence would be generated in the flow of air in the upper part of airway. In CFD, $k$–$\varepsilon$ model is the most preferred turbulence model.

Bates *et al*.(12) Studied the dynamics of airflow in short inhalation based on the following boundary condition. The transient velocity at the inlet has been applied to the external hemisphere, whereas the ends of the bronchi extension were kept at a constant pressure. All the surfaces were made to fall under no-slip criteria. No major variations were seen when the radius of the hemisphere defining the external far-field boundary was changed from 0.5 to 1 m from the face center or when the simulation was carried out again by applying a pre-defined pressure as an inlet criteria, instead of describing the velocity. The walls have been assumed to be rigid, as Fodil *et al*. [13] showed that there is not much variation in the area until higher nasal pressure drops are studied, i.e. above 100 Pa.

It can be fairly observed that these condition will vary as per different age, gender, condition of patient, and the focus of study.

**CFD aspect of airflow and particle transport/deposition:**

Air borne particles inhaled have potential to give rise to many diseases which may be as dangerous as lung cancer. The TB (tracheobronchial) deposition values as simulated by CFD by Zhang *et al*.(14) was in accordance with the results obtained analytically. However, significant variations was observed for bifurcation by bifurcation fractions of deposition because of a more realistic inflow conditions and geometric features included in the CFD model.

Saber and Heydari(15)used CFD simulation to predicted the particle deposition fraction of 0.1-10mm under resting activity i.e. 15 l/min and moderate activity i.e. 45 l/min breathing patterns in order to compare with the results predicted by Martonen *et al*.(16) at the same flow rate. The velocity field and fraction of deposition of the particles in the human trachea and in the first and third generations of the tracheobronchial tree were examined in their study along with the CFD simulation. The flow of air into the trachea was assumed to be in the range of 15-60 l/min, as per 4 different levels of activity in an adult male. The fraction of deposition for a particle range of 0.1-10 mm in diameter is found to have a minimum range of 0.1-1mm diameter of the particle, after which it increases for particles of bigger sizes. Also the simulation results, under varied patterns of breathing have proved that the fraction of deposition increases with increase in Reynolds as well as Stokes number.

Kirch *et al*.(17) investigation provides us with a way of describing the particle deposition, sedimentation as well as the clearance of the nano-sized particles within the mucus, with the help of numerical as well as the analytical models. The deposition of the particle as well as its mobility, sedimentation and the clearance were modeled with CFD and represented analytically. In addition to this, grid free CFD methods were used to model the mucus plasticity as a possible path for a complex particle translocation.

Study done by Srivastava *et al*.(18) found that cartilaginous ring's location may lead to deposition of particle locally and injury to wall. Inspiratory as well as expiratory flow of air was studied on the basis of velocity, static pressure and wall shear stress distribution. In addition to this, inspiratory two phase flow was examined for particle deposition and the numerical values for velocity, pressure and wall shear stress can be determined at any time in the model. Their study confirmed that valuable information for clinical diagnosis can be effectively provided by CFD. They also concluded that this study of particle deposition can play a vital role in prediction of toxic particle deposition and drug delivery information.

The primary objective of Gorgi *et al.* (4) in their investigation was researching the behavior of the airflow and the fraction of particle deposition in varied conditions of breathing, like, light breathing condition i.e. 15 l/min, normal breathing condition i.e. 30 l/min and heavy breathing condition i.e. 60 l/min. A clear picture of air particle dynamics for varying breathing condition in TB airways was obtained which is very helpful for the medical expert to understand the drug delivery and also for medication purpose.

CFD can also play a vital role in understanding the effect of Tracheomalacia in some children where treatment is difficult in clinical condition as the current method is not suitable for them.

**CFD aspect in drug delivery and aerosol targeting:**

Braker *et al*. (19) used this multidisciplinary CFD tool in asthamatic patients to find the consequence of bronchodilator and they shown that CFD is capable to descry variation in airway resistance and changes in the spirometric values only partially describe the effect of a bronchodilator in asthmatics, since no information regarding the regional variations can be obtained. Imaging methods like high-resolution computed tomography (HRCT) contribute additional information but lack precise and comprehensive on airway responses specifically. The objective of their study was to improvise the present imaging techniques by consecutive analysis of the imaging data using CFD. All the variations in CFD based calculated airway resistance are majorly in accordance with the visible changes in the spirometric values and thus, this validates the CFD method for the applications that have been examined.

According to Ma and Lutchen (20) Quantitative data on the deposition of aerosol particles in the respiratory tract of human beings are essential for getting to know the reasons for the occurrence of certain types of lung diseases and for efficiently modeling drug delivery systems through inhalation.The deposition of aerosols in a 3-D anatomically based human large-medium airway was modeled using CFD in their study. The majority deposition in the large-medium airway model showed that the overall TB deposition were controlled by the large-medium airways for micrometer-sized aerosol particles. The methods and data developed in their study, has given useful ways of simulation of deposition of aerosol in the human lungs on the basis of realistic geometry of the lungs.

Inthavong *et al*.(21) study examines two methods of aerosol medication delivery, firstly a perfect case for drug particle delivery under normal conditions of the breathing cycle and  secondly  for a greater effort during the inhalation along with hold of the breath. Profiles of velocity, local fractions of deposition and patterns of deposition of the aerosol particles in the first 6 generations of an airway model were represented in the simulation by them. The velocity profiles were also utilized to determine the pattern of deposition formed. It was then observed that a deep inhalation with a breath hold of around 2 seconds did not essentially imply greater deposition up to the sixth branch generation, but instead, there was an increase in the deposition in the first few airway generations. Also, the breath hold provides for a deposition by sedimentation which aids in targeted deposition locally. Visual representation of the deposition of particles showed local 'hot-spots' in areas where particle deposition was significant in the airway of the lungs.

Xi *et al* (22) did a commendable research which numerically analyses the age-related effects on the flow of air and dynamics of the aerosol particles in image based nose-throat models of a 10-day-old infant, a 7-month-old baby, a 5-year-old child, and a 53-year-old adult. Variations in airway physiology, breathing resistance and aerosol filtering efficiency among the four models were quantified and comparisons were made by them. A high fidelity fluid transport particle model was used for the sake of simulating the multi-regime airflows and the particle transport within the nasal-laryngeal airways and ultrafine particles were studied under breathing conditions varying from sedentary to heavy activities. Results obtained in their study show that the nasal–laryngeal airways at different ages, albeit differ significantly in morphology and dimension, do not significantly affect the total deposition fractions or maximum local deposition enhancement for ultrafine aerosols. Moreover, the particle deposition in the sub-regions of interest is not the same for the four models considered.

Technologies of drug delivery are considered an important and essential area in the field of biomedicine. The effective drug delivery is when active agents are supplied only to the specific area in the body thereby reducing harmful effects. Therefore, an arithmetic analysis of the magnetic drug targeting (MDT) through tracheobronchial airways with the use of aerosol drugs names polystyrene particle (PMS040) in the lungs of a human was shown assuming one-way coupling on the movement and capture of the magnetic particle was observed by Pourmehran *et al*.(23) . A realistic 3-D geometry based on the images of the CT scan was given for CFD simulation. As per the results, the magnetic field enhances the efficiency of deposition of the particles within a specified target region. Also, the deposition efficiency and MDT technique has a direct relation with a greater particle diameter, for a magnetic number of 1 Tesla (T) and below. In addition to this, it can also be observed that there exists an inverse relationship between the diameter of the particle and the efficiency of deposition when the magnetic number is greater than 1T.

CFD allows us to determine the deposition of pharmaceutical aerosols was observed from the study done by Tian *et al*.(24). Earlier investigations have not differentiated the CFD results of deposition all through the lungs and the in vivo data for the monodisperse aerosol. The predictions of CFD of the total lung deposition was in accordance with the in vivo data, thus giving a relative error of 6%, and averaged over aerosol sizes of 1-7 μm. In the case of dry powder inhaler and softmist inhaler, the deposition was studied by them in the mouth- throat region, the central airways, and the intermediate plus peripheral airways .CFD results showed a mean relative error of less than 10%, for each inhaler across those areas. Simulations performed using CFD with the stochastic individual pathway approach to modeling were proven to exactly determine the regional deposition across the lungs for multiple types of aerosol and also for different in vivo analysis methods.

.

**CFD aspect in stent implementation:**

The method of CFD, which gives an approximation of the pressure drop in the airway, before and after the inclusion of the stent, was proposed in the research of Ho *et al*.(3). The method of finite volume was used for this purpose. The pressure field was determined using the Navier Stokes equation. The prescribed method was examined for seven healthy people i.e. the control group, and in 14 patients, who were allotted into two groups – tracheal stenosis and bronchial stenosis. The results obtained proved that the drop in the pressure after the implementation of the tracheal stent became significantly lesser. In the case of implementation of the bronchial stent, the airway resistance increased, however, insignificantly. A complete model was provided by authors in order to methodically assess patients who have obstructive lung tumor. With the help of the limited examples, they observed that the numerically obtained results using CFD, correlates well with the results of the PFT's, thus implying that CFD can be a clinical prognosis tool to examine lung cancer patients who have poor general condition and for whom PFTs cannot be efficiently applied to. Moreover, they have also observed that TB (Tracheobrochial) stent placement indeed improves the quality of life of the cancer patients by reducing the resistance in their respiratory tract. The CFD model in their research was thus extremely helpful in not only assessing the outcome of personalized placement of stent, but also following up of every patient who goes through the process of tracheal and/or bronchial stent placement.

The CFD model of a healthy, a stenotic as well as a post-operatory stented human trachea was generated by Malve *et al*.(25)in order to study the mechanism of respiration under the prescribed physiological boundary conditions. In order to achieve that, waveforms of the outlet pressure were obtained from patient specific spirometries using a method that makes it possible to compute the peripheral impedance of the truncated bronchial generation, simulating the lungs as fractal networks. The suggested impedance method presents an agreeable tool in order to determine the physiological pressure conditions that are otherwise impossible to obtain *in vivo*. There methodology can also be used on healthy, pre and post operatory tracheas representing the possibility of finding out, via numerical modeling, the pressure field as well as the flow, before and after the operation

CFD as a tool, provides for the visualization and comprehension of the flow physics involved with stent implantation. Shih *et al*.(9) research uses the patients CT scan images in order to examine the correlation between pressure drop and flow rate of air in the tracheal airway. Additionally, the cross-sectional areas of the patient's airway were calculated and the area stenosis ratios were obtained. The numerical results obtained using CFD exhibit the possibility of evaluating airway resistance on the basis of the CT scan images. In clinical practice, it is generally not easy to examine patients who have severe airway obstruction using the symptoms score or a pulmonary function test as a result of their critical condition. The suggested non-invasive method for finding out the airway obstruction severity aids in helping patients who cannot do pulmonary function tests. Zhu *et al*.(26) have examined a patient-specific model of LSCTS (Long segment congenital tracheal stenosis) with complete tracheal rings and bridging bronchus (BB). To improve therapies such as LSCTS with distal bronchus stenosis, local aerodynamics information is very limited. In their study CFD was used to study the local aerodynamics around BB and after tracheal surgery in inspiratory phase as well as the expiratory phase. Average pressure drop, wall shear stress, streamlines and loss of energy were determined in order to evaluate the outcome of the surgeries. To analyze the local aerodynamics is essential for the improvement of surgical therapies of the LSCTS.

**Conclusion:**

From the capabilities of CFD discussed above, it can predict almost real picture of airflow, particle deposition/transportation, drug delivery etc. Also, from respiratory point of view, it has made considerable progress to aid in analyzing tracheal stent, tracheomalacia, etc. CFD can be seen as one of the promising tool which can be used at various stage starting from diagnosis to final treatment which may include surgery done with the help of computers. As seen from the research done, cases in which patients who cannot perform PFT, exact result as accurate as PFT can be obtained with the help of CFD. Specific research is necessary to bridge the gap so that CFD can be directly used by doctors as a handy clinical tool. CFD can simplify and assists many flow related problems which are still complex and difficult to solve in respiratory science.

**Reference :**

1. Kleinstreuer C and Zhang Z . Airflow and particle transport in the human respiratory system. Annu. Rev. Fluid Mech. 2010; 42: 301-34.

2. Walters DK, Burgreen GW, Lavallee DM, Thompson DS, and Hester RL. Efficient, Physiologically Realistic Lung Airflow Simulations. IEEE Transactions on Biomedical Engineering. 2011; 58(10) :3016-9.

3. HO CY,  Liao HM, Tu CY,Huang CY, Shih CM, Su MYL, Chen JH and Shih TC. Numerical analysis of airflow alteration in central airways following tracheobronchial stent placement. Experimental Hematology & Oncology. 2012; 1:23.

4. Gorji MR, Pourmehran O, Bandpy MG, Gorji TB. CFD simulation of airflow behavior and particle transport and deposition in different breathing conditions through the realistic model of human airways. Journal of Molecular Liquids. 2015; 209:121-33.

5. Pu J, Fuhrman C, Good WF, Sciurba FC , Gur D. A Differential Geometric Approach to Automated Segmentation of Human Airway Tree. IEEE Transactions On Medical Imaging. 2011; 30(2); 266-78.

6. Dong S, Xiangjun S , Fusheng L, Mintang L, Xinxi X. Creation of standardized geometrical model of human upper respiratory airways. IEEE(978-1-4673-5887-3/13)**.** 2013; 38-42.

7. Johari NH, Osman K, Helmi NHN, Mohammed A. Kadir RA. Comparative analysis of realistic CT-scan and simplified human airway models in airflow simulation. Computer Methods in Biomechanics and Biomedical Engineering.2015; 18(1): 48-56.

8. *Nardelli P, Khan KA, Corvo A, Moore N, Murphy MJ, Twomey M,Connor OJO, Kennedy MP, Estepar RSJ , Maher MM, Murphy PC. Optimizing parameters of an open-source airway segmentation algorithm using different CT images. BioMed Eng OnLine. 2015; 14:62.

9. Shih TC, Hsiao HD, Chen PY, Tu CY, Tseng TI, Ho YJ.   Study of Pre-and Post-Stent Implantation inthe Trachea Using Computational Fluid Dynamics. Journal of Medical and Biological Engineering. 2014; 34(2):150-6.

10. Nithiarasu P, Hassan O, Morgan K, Weatherill NP, Fielder C, Whittet H, Ebden P, Lewis KR. Steady flow through a realistic human upper airway geometry. Int. J. Numer. Meth. Fluids.  2008; 57:631–51.

11. Wang Y, Liu Y, Sun X, Yu S, Gao F. Numerical analysis of respiratory flow patterns within human upper airway. Acta Mech Sin. 2009; 25:737– 46.

12. Bates AJ, Doorly DJ, Cetto R, Calmet H, Gambaruto AM, Tolley NS, Houzeaux G, Schroter RC. Dynamics of airflow in a short inhalation. J. R. Soc. Interface. 2015; 12:20140880.

13. Fodil R, Brugel-Ribere L, Croce C *et al*. Inspiratory flow in the nose: a model coupling flow and vasoerectile tissue distensibility. J Appl Physiol. 2005;98:288–95

14. Zhang Z,  Kleinstreuer C, KIM C. Airflow and Nanoparticle Deposition in a 16-Generation Tracheobronchial Airway Model. Annals of Biomedical Engineering. 2008; 36(12) : 2095– 110.

15. Saber EM, Heydari G. Flow patterns and deposition fraction of particles in the range of 0.1–10 mm at trachea and the first third generations under different breathing conditions. Computers in Biology and Medicine. 2012; 42: 631–38.

16. Martonen TB, Musante CJ, Segal RA, Schroeter JD, Hwang D, Dolovich MA, Burton R, Spencer RM, Fleming JS.Lung models: strengths and limitations. Respir. Care. 2000; 45: 712–36.

17. Kirch J, Guenther M, Schaefer UF ,Schneider M,  Lehr CM. Computational fluid dynamics of nanoparticle disposition in the airways: mucus interactions and mucociliary clearance. Comput Visual Sci .2011; 14:301–08.

18. Srivastav VK, Paul AR, Jain A. Effects of cartilaginous rings on airflow and particle transport through simplified and realistic models of human upper respiratory tracts. Acta Mechanica Sinica. 2013; 29(6):883–92.

19. Backera JWD, Vosa WG, Devoldera A, Verhulsta SL, Germonpre P, Wuytsb FL, Parizelc PM, Backer WD. Computational fluid dynamics can detect changes in airway resistance in asthmatics after acute bronchodilation. Journal of Biomechanics .2008; 41: 106–13.

20. Ma B and Lutchen KR. CFD Simulation of Aerosol Deposition in an Anatomically Based Human Large–Medium Airway Model. Annals of Biomedical Engineering. 2009; 37(2):271–85.

21. Inthavong K, Choi LT, Tu J, Ding S, Thien F. Micron particle deposition in a tracheobronchial airway model under different breathing conditions. Medical Engineering & Physics. 2010; 32: 1198–212.

22. Xi J, Berlinski A, Zhou Y, Greenberg B, Ou X. Breathing Resistance and Ultrafine Particle Deposition in Nasal–Laryngeal Airways of a Newborn, an Infant, a Child, and an Adult. Annals of Biomedical Engineering. 2012; 40(12): 2579–95.

23. Pourmehran O ,Gorji MR, Bandpy MG, Gorji TB. Simulation of magnetic drug targeting through tracheobronchial air- ways in the presence of an external non-uniform magnetic field using Lagrangian magnetic particle tracking. Journal of Magnetism and Magnetic Materials. 2015; 393:380–93.

24. Tian G, Hindle M, Lee S, Longest PW. Validating CFD Predictions of Pharmaceutical Aerosol Deposition with In Vivo Data. Pharm Res. 2015; 32:3170–87.

25. Malve M, Chandrac S, Villalobosd JLLP, Finolc EA, Gineld A ,Doblare A. CFD analysis of the human airways under impedance-based boundary conditions: application to healthy, diseased and stented trachea. Computer Methods in Biomechanics and Biomedical Engineering. 2013; 16(2):198-216.

26. Zhu L, Liu J, Zhang W, Sun Q,Hong H, Du Z, Liu J, QianY, Wang Q, Umezu M. Computational Aerodynamics of Long Segment Congenital Tracheal Stenosis with Bridging Bronchus. IEEE (978-1-4799-7862-5/13). 2015.